# Reaction Mechanism of the Selective Reduction of $CO_2$ to CO by a Tetraaza $[Co^{II}N_4H]^{2+}$ Complex in the Presence of Protons

Alejandro J. Garza
*Joint Center for Artificial Photosynthesis, Lawrence Berkeley National Laboratory, Berkeley, CA 94720, USA*

Srimanta Pakhira
*Department of Physics, Scientific Computing, Materials Science and Engineering,*
*High Performance Material Institute, Condensed Matter Theory-National High Magnetic Field Laboratory,*
*Florida State University, Tallahassee, FL 32310, USA and*
*Department of Chemical & Biomedical Engineering,*
*Florida A&M University - Florida State University,*
*Joint College of Engineering, Tallahassee, FL 32310, USA*

Alexis T. Bell
*Joint Center for Artificial Photosynthesis, Lawrence Berkeley National Laboratory, Berkeley, CA 94720, USA and*
*Department of Chemical and Biomolecular Engineering,*
*University of California, Berkeley, CA 94720, USA*

Jose L. Mendoza-Cortes[*]
*Joint Center for Artificial Photosynthesis, Lawrence Berkeley National Laboratory, Berkeley, CA 94720, USA*
*Department of Physics, Scientific Computing, Materials Science and Engineering,*
*High Performance Material Institute, Condensed Matter Theory-National High Magnetic Field Laboratory,*
*Florida State University, Tallahassee, FL 32310, USA and*
*Department of Chemical & Biomedical Engineering,*
*Florida A&M University - Florida State University,*
*Joint College of Engineering, Tallahassee, FL 32310, USA*

Martin Head-Gordon[†]
*Joint Center for Artificial Photosynthesis, Lawrence Berkeley National Laboratory, Berkeley, CA 94720, USA and*
*Department of Chemistry, University of California, Berkeley, CA 94720, USA*

The tetraaza $[Co^{II}N_4H]^{2+}$ complex (**1**) is remarkable for its ability to selectively reduce $CO_2$ to CO with 45% Faradaic efficiency and a CO to $H_2$ ratio of 3:2. We employ density functional theory (DFT) to determine the reasons behind the unusual catalytic properties of **1** and the most likely mechanism for $CO_2$ reduction. The selectivity for $CO_2$ over proton reduction is explained by analyzing the catalyst's affinity for the possible ligands present under typical reaction conditions: acetonitrile, water, $CO_2$, and bicarbonate. After reduction of the catalyst by two electrons, formation of $[Co^IN_4H]^+-CO_2^-$ is strongly favored. Based on thermodynamic and kinetic data, we establish that the only likely route for producing CO from here consists of a protonation step to yield $[Co^IN_4H]^+-CO_2H$, followed by reaction with $CO_2$ to form $[Co^{II}N_4H]^{2+}-CO$ and bicarbonate. This conclusion corroborates the idea of a direct role of $CO_2$ as a Lewis acid to assist in C−O bond dissociation, a conjecture put forward by other authors to explain recent experimental observations. The pathway to formic acid is predicted to be forbidden by high activation barriers, in accordance with the products that are known to be generated by **1**. Calculated physical observables such as standard reduction potentials and the turnover frequency for our proposed catalytic cycle are in agreement with available experimental data reported in the literature. The mechanism also makes a prediction that may be experimentally verified: that the rate of CO formation should increase linearly with the partial pressure of $CO_2$.

## I. INTRODUCTION

Transition metals supported by tetradentate, redox-non-innocent ligand macrocycles are well known for reducing protons to dihydrogen [1–8]. However, some are also capable of reducing $CO_2$ to CO, formate, and oxalate [9–15]. This offers the intriguing possibility of utilizing $CO_2$ to generate valuable chemicals or store energy from wind or solar radiation in the form of liquid fuels. Selectivity towards generating products of $CO_2$ reduction is one of the main limitations for practical applications of this process, although there has been steady progress in this area over the last decades.

$CO_2$-reducing cobalt and nickel phthalocyanines were first reported by Meshitsuka *et al.* in

---

[*]Electronic address: mendoza@eng.famu.fsu.edu
[†]Electronic address: mhg@cchem.berkeley.edu

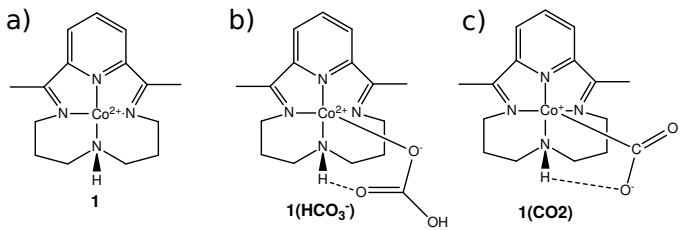

FIG. 1: Structure of (a) $[Co^{II}N_4H]^{2+}$ (**1**), (b) adduct of **1** with $HCO_3^-$ (c) adduct of (two-electron reduced) **1** with $CO_2$

1974 [9]. Since then, a wide-range of catalysts based on different host ligands have been reported: cyclam [16–18], phthalocyanines [19–21], porphyrins [22–24], polypyridines [25–27], to name a few. More recently, it has been demonstrated that tetraaza macrocyclic complexes of late first row transition metals, such as $[Co^{II}-N_4H(MeCN)]^{2+}[B(C_6H_5)_4^-]_2$ ($N_4H$ =2,12-dimethyl-3,7,11,17-tetraazabicyclo-[11,3,1]-heptadeca-1(17),2,11,13,15-pentaene), are efficient catalysts for hydrogen evolution [28–30] and $CO_2$ reduction [31–33], either electrocatalytically, or by using a visible light sensitizer in the presence of a sacrificial electron donor. Among the tetraaza catalysts, $[Co^{II}N_4H]^{2+}$ (compound **1** in Figure 1) stands out for its ability to generate CO with 45% Faradaic efficiency and an unusual CO to $H_2$ ratio of 3:2 near the $Co^{I/0}$ potential [30]. This catalyst is remarkable because cobalt $N_4$-macrocycles are known for favoring proton reduction [34]. Furthermore, **1** is highly selective towards CO over other carbon products: no formate or oxalate are produced [30]. The cause of the atypical catalytic activity of **1** is not understood, neither is the mechanism by which **1** reduces $CO_2$, although **1(HCO$_3^-$)** and **1(CO$_2$)** (see Figure 1) have been observed by FT-IR spectroscopy [35]. The presence of **1(HCO$_3^-$)** indicates that bicarbonate may be involved in the reaction mechanism; however, this is unusual for this type of compound and has not been confirmed either experimentally or theoretically.

Herein, we employ density functional theory (DFT) to explain the selectivity of **1** towards reduction of $CO_2$ to CO. Our calculations corroborate the preference of **1** for binding $CO_2$ rather than protons, and provide insight into the structural origin of this preference. We derive the most likely mechanism for $CO_2$ reduction, and compute observables such as the turnover frequency (TOF) using the energetic span model. These results agree with, and explain, various notable experimental observations, including the selectivity for CO over formate. Importantly, our conclusions strongly support the hypothesis that bicarbonate plays a critical role in the reaction mechanism. This study is an extension of our previous theory-experiment work on the subject [36].

## II. THEORY AND COMPUTATIONAL DETAILS

Gas phase calculations were carried out using the QChem software package [37]. Calculations in water and acetonitrile were performed with the polarizable continuum model [38] implemented in Gaussian [39] (experiments on **1** are normally carried out on wet acetonitrile, i.e. ≈ 10 M water in $CH_3CN$). The results that we report here are for the acetonitrile medium, but we remark that reaction energies in water and acetonitrile do not differ significantly. Transition and ground state geometries were optimized at the BP86/6-31G(d,p) level using spin-unrestricted Kohn–Sham determinants. This level of theory was selected on the basis of good agreement with available experimental structural and spin-state data for **1**; BP86 yielded better results than PBE, B3LYP, and B97X-D (further details were presented in our previous work on ref. [36]). The results reported here are for low-spin states of the catalyst, which corresponds to the ground state according to both BP86 simulations and experimental measurements [30, 36]. Free energies were obtained by correcting electronic energies with their corresponding enthalpic and entropic contributions at 298 K using vibrational frequency data and standard formulas [40]. We noticed, however, that the ideal gas, rigid rotor and harmonic oscillator formulas significantly overestimate the entropy of $CO_2$ and $H_2O$ in solution (translation and rotation motions are sharply reduced in solution [41, 42]). This results in spurious $\Delta G$ values for bimolecular reactions. Thus, we used reported experimental entropies in aqueous solution to compute the thermal corrections of these molecules ($\Delta S_{sol \to gas}$ = 126.0 and 118.9 J/mol·K for $CO_2$ and $H_2O$, respectively [43–45]).

We also encountered difficulties in calculating free energies for protonation reactions because of the poor description of the solvation energy of protons by implicit solvent models. To circumvent this issue, we followed other authors [46] and employed accurate proton solvation energies from high-level *ab initio* calculations (−260.2 kcal/mol for acetonitrile [47]) which agree, within experimental error, with experimental values.

Turnover frequencies (TOFs) for the possible catalytic cycles were estimated with the energetic span model [48–50]:

$$\text{TOF} = \frac{kT}{h} \cdot \frac{e^{-\Delta G_r/kT} - 1}{\sum_{i,j} e^{(T_i - I_j - \delta G_{ij})/kT}} \quad (1)$$

$$\delta G_{ij} = \begin{cases} \Delta G_r, & \text{if } i > j \\ 0, & \text{if } i \leq j, \end{cases}$$

where $\Delta G_r$ is the total reaction free energy, and $T_j$ and $I_j$ are the free energies of the $j$th transition state and intermediate in the cycle, respectively.

TABLE I: Calculated standard reduction potentials (*vs* the SHE) and free energies of binding (in acetonitrile) for complexes of **1** with bicarbonate, acetonitrile, water, and carbon dioxide in different oxidation states.

| Species | Ligand ($E°$/V) | | | | |
|---|---|---|---|---|---|
| | None | $HCO_3^-$ | MeCN | $H_2O$ | $CO_2$ |
| $[Co^{II}N_4H]^{2+}$ | +0.08 | −0.71 | −0.22 | +0.07 | +0.47 |
| $[Co^{I}N_4H]^{+}$ | −1.33 | −1.69 | −1.42 | −1.32 | −1.01 |
| | $\Delta G_{bind}$/kcal·mol$^{-1}$ | | | | |
| $[Co^{II}N_4H]^{2+}$ | | −28.7 | −7.7 | −9.8 | +2.9 |
| $[Co^{I}N_4H]^{+}$ | | −10.3 | +0.9 | −6.2 | −6.1 |
| $[Co^{0}N_4H]$ | | −1.9 | +4.1 | −6.4 | −13.3 |

## III. RESULTS AND DISCUSSION

### A. Binding to water, bicarbonate, and carbon dioxide

Table I lists the acetonitrile-medium free energies for binding **1** to species that would normally be found in the reaction: acetonitrile, water, bicarbonate, and carbon dioxide. Three oxidation states of **1** are considered; reduction potentials *vs* the standard hydrogen electrode (SHE) are also reported (the absolute SHE potential is taken as 4.43 V [51]). In its highest oxidation state, $[Co^{II}N_4H]^{2+}$ is unable to bind $CO_2$ ($\Delta G_{bind}$ = 2.9 kcal/mol), and instead binds preferentially to bicarbonate ($\Delta G_{bind}$ = −28.7 kcal/mol). Positive charge on the metal center and a favorable O···HN hydrogen bond (see Fig. 1b) are responsible for the strong interaction between $HCO_3^-$ and **1**; the Mulliken charge on Co reduces from 1.12 to 0.97 upon complexation. The hydrogen bond between one of the oxygen atoms in bicarbonate and the NH group in **1** has been observed experimentally by infrared spectroscopy [35], and is present in the BP86-optimized geometry; the length and angle of the H-bond are 1.703 Å and 163.6°, respectively, indicating a strong interaction. Consistent with these results, **1(HCO$_3^-$)** has been experimentally determined to be the dominant species for the oxidized catalyst in the presence of $CO_2$ in wet acetonitrile [35]. The calculated $E°$ = −0.22 V for the $[Co^{II}N_4H]^{2+}$–MeCN complex is very close to the measured reduction potential potential of $[Co^{II}N_4H]^{2+}$ in acetonitrile [30] ($E°$ = −0.28 V).

Upon reduction of the catalyst by one electron, $CO_2$ is calculated to have substantial binding with the catalyst ($\Delta G_{bind}$ = −6.1 kcal/mol), but $HCO_3^-$ is still largely the preferred ligand ($\Delta G_{bind}$ = −10.3 kcal/mol). The affinity for $H_2O$ is anticipated to be similar to that for $CO_2$ ($\Delta G_{bind}$ = −6.2 kcal/mol). These findings are in accordance with the spectroscopic measurements in ref. [52], which detected $[Co^{II}N_4H]^{2+}$–$HCO_3^-$, $[Co^{I}N_4H]^{+}$–$HCO_3^-$, and $[Co^{I}N_4H]^{+}$–$CO_2$, but not $[Co^{II}N_4H]^{2+}$–$CO_2$. However, the $[Co^{II}N_4H]^{2+}$–$H_2O$ cannot be deconvoluted in the FT-IR measursments as it is covered under the large bulk water signal. The calculated standard reduction potential of $[Co^{I}N_4H]^{+}$ ($E°$ = −1.33 V *vs* the SHE) is also in agreement with the reported experimental value ($E°$ = −1.24 V) in MeCN [30]. The $E°$ = −1.42 V for $[Co^{I}N_4H]^{+}$–MeCN is also reasonably close to experiment.

It is not until **1** is reduced by a second electron that $CO_2$ binding ($\Delta G_{bind}$ = −13.3 kcal/mol) becomes dominant over binding to water (−6.4 kcal/mol) and bicarbonate (−1.9 kcal/mol). The doubly-reduced $[Co^{0}N_4H]$ is, however, the catalytically active species: reduction products are not observed unless a sensitizer with enough reduction potential to generate $[Co^{0}N_4H]$ is present [35, 36]. Hence, the unusual selectivity of **1** for $CO_2$ reduction can be traced to the greater affinity of $[Co^{0}N_4H]$ for $CO_2$ over water. This enhanced $CO_2$ affinity can in turn be attributed to two factors: one electronic and one structural. The first one is related to a charge transfer from the catalyst to $CO_2$: when bound to the doubly reduced neutral catalyst, $CO_2$ becomes partially negative (resulting in a bent structure with a bond angle of 132.0°), whilst the catalyst acquires a partial positive charge localized at the transition metal center. This picture is corroborated by the the sum of the Mulliken charges in the oxygen atoms of $CO_2$ (−1.0), and the charges on cobalt (0.66), and the carbon atom of $CO_2$ (0.39). The structural factor is a hydrogen bond between a $CO_2$ oxygen atom and the NH group in the tetraaza ring; the length and angle of this H-bond are 1.797 Å and 139.6°, respectively. In comparison, the length of this same H-bond is 1.911 Å in the singly reduced $[Co^{I}N_4H]^{+}$–$CO_2$ complex. Thus, the (Lewis) acid-base properties of the catalyst synergize with its structure to boost $CO_2$ affinity by transferring negative charge to the latter and, consequently, strengthening the O···HN bond. Note that although intramolecular H-bonds are rare in protic solvents [53, 54], they can be quite favorable in aprotic solvents and, in fact, signals corresponding to a $CO_2$···HN H-bond are present in the IR spectra of the reduced catalyst in acetonitrile [35].

### B. Mechanism of $CO_2$ reduction

The most likely mechanism of $CO_2$ reduction is shown in Fig. 2. The catalyst remains preferentially bound to $HCO_3^-$ in the +2 and +1 oxidation states. Complexation with bicarbonate lowers the reduction potential of of the catalyst by −0.79 and −0.36 V for the first and second electron additions, respectively. However, as noted above, $CO_2$ binding is thermodynamically favored over $HCO_3^-$ binding for the doubly-reduced, neutral catalyst by 11.4 kcal/mol. As detailed below and corroborating experimental observations, we find that the complex that can roughly be described as $[Co^{I}N_4H]^{+}$–$CO_2^-$ (**5** in Fig. 2) is the key intermediate leading to CO.

In our derivation of the mechanism below, we assume that protons are available to facilitate reactions. This assumption is reasonable because spectroscopic measurements [35] and our own calculations in Table I indicate that, in a saturated CO$_2$ solution, the oxidized catalyst exists predominantly as **1(HCO$_3^-$)**. The bicarbonate ligand is generated exclusively from CO$_2$ and H$_2$O, resulting in protons being available in the wet acetonitrile medium. Following other authors [46], the total free energy of a proton in solution is determined from high-level *ab initio* calculations [47] and utilized to compute free energy differences (see Theory and Computational Details).

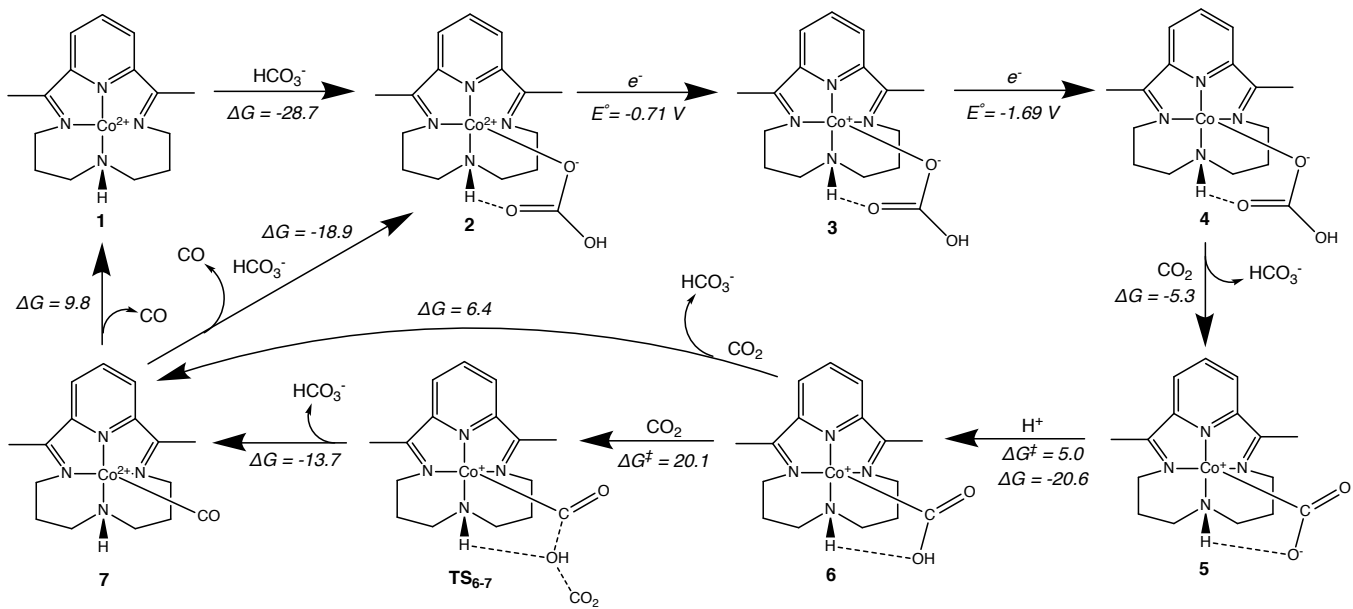

FIG. 2: Proposed mechanism for the CO$_2$ reduction catalytic cycle. Free energies are in kcal/mol.

After formation of [Co$^{\rm I}$N$_4$H]$^+$–CO$_2^-$, the most favorable elementary step that we have identified is protonation of the CO$_2$ oxygen atom:

$$[\text{Co}^{\rm I}\text{N}_4\text{H}]^+ - \text{CO}_2^- + \text{H}^+ \rightarrow [\text{Co}^{\rm I}\text{N}_4\text{H}]^+ - \text{CO}_2\text{H}. \quad (2)$$

This is a fast and highly exergonic process with a calculated $\Delta G^\ddagger = 5.0$ kcal/mol and a $\Delta G = -20.6$ kcal/mol. Alternative reactions such as protonation on the C atom

$$[\text{Co}^{\rm I}\text{N}_4\text{H}]^+ - \text{CO}_2^- + \text{H}^+ \rightarrow [\text{Co}^{\rm I}\text{N}_4\text{H}]^+ - \text{HCO}_2 \quad (3)$$

or reaction with CO$_2$ to form CO$_3^{2-}$ and CO

$$[\text{Co}^{\rm I}\text{N}_4\text{H}]^+ - \text{CO}_2^- + \text{CO}_2 \rightarrow [\text{Co}^{\rm II}\text{N}_4\text{H}]^{2+} - \text{CO} + \text{CO}_3^{2-} \quad (4)$$

do not appear to be competitive. The reaction in eq. 3 would (most likely) lead to formate, which is not produced by catalyst **1**. Not only this, but protonation on the carbon atom is chemically counterintuitive given that the negative charge is localized in the O atoms. Consistent with this notion, the free energy barrier for this reaction is estimated to be much higher ($\approx 32$ kcal/mol; *vide infra*) than the 5 kcal/mol for oxygen protonation. Likewise, the $\Delta G^\ddagger$ to generate CO and carbonate ion (eq. 4) is predicted to be about 40 kcal/mol. This explains the experimental observation of CO$_2$ reduction not taking place in the absence of water [36]: protons coming from the reaction of CO$_2$ with water are required for eq. 2 to occur and eventually lead to CO, as the proton-free eq. 4 is not a viable path. In [Co$^{\rm I}$N$_4$H]$^+$–CO$_2^-$, the C–O bond has double bond character: the length of the C–O bond for the oxygen forming the H-bond with the NH group is 1.265 Å, whereas the length of the other C–O bond is 1.240 Å. This is, presumably, the reason for the high $\Delta G^\ddagger$ required to break one of these bonds via eq. 4.

After the initial protonation, eq. 2, the lowest energy path consists of the CO$_2$H group reacting with CO$_2$ to form bicarbonate and CO:

$$[\text{Co}^{\rm I}\text{N}_4\text{H}]^+ - \text{CO}_2\text{H} + \text{CO}_2 \rightarrow [\text{Co}^{\rm II}\text{N}_4\text{H}]^{2+} - \text{CO} + \text{HCO}_3^-. \quad (5)$$

With a calculated $\Delta G^\ddagger = 20.1$ kcal/mol, this is the rate-determining step of the catalytic cycle. Nonetheless, the activation free energies for other possible reactions are much higher in our simulations (see discussion below and SI). The barrier for a second protonation on an oxygen atom to form CO and water

$$[\text{Co}^{\rm I}\text{N}_4\text{H}]^+ - \text{CO}_2\text{H} + \text{H}^+ \rightarrow [\text{Co}^{\rm II}\text{N}_4\text{H}]^{2+} - \text{CO} + \text{H}_2\text{O} \quad (6)$$

was computed as $\Delta G^\ddagger = 61.6$ kcal/mol. It is not immediately clear why the barrier to form water is so high as compared to that for forming bicarbonate; we verified various possible ways in which the proton could attack CO$_2$H but did not find lower activation energies. We conclude that the CO$_2$H group attached to the catalyst is an acid, not a base, and, in a certain way, behaves similarly to bicarbonate or carbonic acid. An additional factor contributing to the HCO$_3^-$ preference is the electrostatic interaction between the positively-charged catalyst and the negatively-charged bicarbonate. As reflected on Table I, the interaction of $HCO_3^-$ with the doubly charged catalyst is about 20 kcal/mol stronger then the interaction of the latter with water.

Two closely related steps that can occur are protonation of the OH end of $CO_2H$ without simultaneous C–O bond cleavage

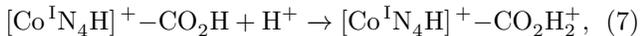
$$[Co^I N_4 H]^+ - CO_2 H + H^+ \rightarrow [Co^I N_4 H]^+ - CO_2 H_2^+, \quad (7)$$

and reaction of $[Co^I N_4 H]^+ - CO_2 H$ with water to form CO, $OH^-$, and water

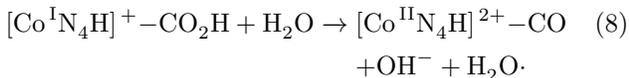
$$[Co^I N_4 H]^+ - CO_2 H + H_2 O \rightarrow [Co^{II} N_4 H]^{2+} - CO \quad (8)$$
$$+ OH^- + H_2 O.$$

However, the structure of $[Co^I N_4 H]^+ - CO_2 H_2^+$ does not appear to be stable; we were unable to optimize a geometry for this complex that did not revert to $[Co^{II} N_4 H]^{2+} - CO$ and water. Eq. 8 is analogous to eq. 6, except that water, instead of a proton, is acting as a Brønsted acid. This would increase $\Delta G^\ddagger$ compared to the already too-high $\Delta G^\ddagger$ of eq. 6. Therefore we do not consider the reaction in eq. 8 a possibility.

Another plausible reaction that $[Co^I N_4 H]^+ - CO_2 H$ could undergo consists of a proton attack on the C atom to yield formic acid:

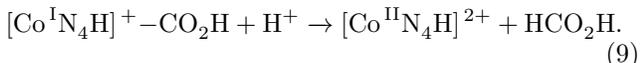
$$[Co^I N_4 H]^+ - CO_2 H + H^+ \rightarrow [Co^{II} N_4 H]^{2+} + HCO_2 H. \quad (9)$$

The BP86 functional gives a $\Delta G^\ddagger = 31.8$ kcal/mol for this process. This is again in agreement with chemical intuition as the C atom has a partial positive charge (Mulliken charge = 0.35), and one would expect it to be less basic than the O atoms. But there seem to be further reasons related to the catalyst's structure that are responsible for the selectivity towards CO as a product. When we optimize the structure of the transition state, the proton is first captured by the negatively-charged N atoms of the tetraaza ring, before being transferred to the carbon atom. This appears to be a result of the orientation of the $CO_2H$ group, which is anchored by the $CO_2H \cdots HN$ hydrogen bond. The basic N atoms in the catalyst do not capture the proton if we allow for the latter to attack from above (rather than from a side/below); however, this results in an even larger barrier ($\Delta G^\ddagger = 60.6$ kcal/mol) because the density that the C atom can donate is located around the C–Co bond. Thus, because of H-bonding, the position of the catalyst's NH group is crucial in determining not only the affinity for $CO_2$ binding, but also the selectivity for CO by preventing a direct proton attack on the C atom to generate formic acid. The fact that the carbon atom does not interact favorably with protons is in line with the aforementioned proposition that the chemistry of the $CO_2H$ group attached to the catalyst is akin to that of carbonic acid or the bicarbonate ion.

Yet another possibility is the protonation of a second oxygen atom in $CO_2H$:

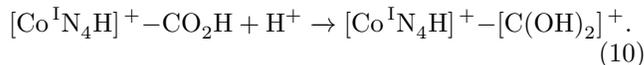
$$[Co^I N_4 H]^+ - CO_2 H + H^+ \rightarrow [Co^I N_4 H]^+ - [C(OH)_2]^+. \quad (10)$$

Eq. 10 is unlikely to occur as its $\Delta G^\ddagger$ is about 30 kcal/mol higher than the $\Delta G^\ddagger$ for CO and bicarbonate production according to eq. 5. The reaction $\Delta G$ for eq. 10 is also substantially endergonic (11.1 kcal/mol). The high relative energy of $[Co^I N_4 H]^+ - [C(OH)_2]^+$ is unsurprising given that it places further positive charge on the C atom that is linked to $Co^+$ center (chemical intuition is, once more, spot-on).

Not only is the mechanism on Fig. 2 predicted to be the main route for $CO_2$ reduction, but the TOF computed for this mechanism is also in agreement with experimental data. Using the energetic span model (eq. 1), we calculate a TOF = $1 \times 10^{-2}$ s$^{-1}$. From experimental data, the TOF for catalyst **1** is about $1 \times 10^{-3}$ s$^{-1}$ [30]. In the energy representation, these rates correspond to effective free energies of activation of 20.1 and 21.4 kcal/mol for the calculated and experimental rates, respectively. Hence, the reaction pathway on Fig. 2—the only pathway in which bicarbonate plays a central role—is the only one compatible with experimental observations, according to our calculations.

As noted earlier, recent experimental studies of catalyst **1** have pointed to bicarbonate as an important player in the catalytic cycle [35]. A direct role of $CO_2$ as a Lewis acid to assist C–O bond dissociation has also been suggested to explain observations related to the activity of a Ni-cyclam catalyst [55]. However, to the best of our knowledge, all experimental observations related to the role of $CO_2/HCO_3^-$ are inconclusive. We are not aware of other theoretical studies considering this possibility, but our results certainly support the ideas put forward in these experimental studies. Based on our calculated $\Delta G^\ddagger$ values for the possible paths to CO, $CO_2$ reduction to CO would not occur if it was not for $CO_2$ itself acting as a Lewis acid to facilitate C–O bond breaking and form $HCO_3^-$. Notice also that bicarbonate plays an additional role by binding to the oxidized catalyst and displacing CO. Otherwise, CO could poison the catalyst ($\Delta G_{bind} = -9.8$ kcal/mol). It is worth noting that a recent (experimental) study has highlighted the importance of bicarbonate in the reduction of $CO_2$ to CO on metallic gold surfaces [56]. This work also demonstrated that the equilibrium reaction

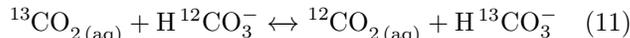
$$^{13}CO_{2\,(aq)} + H^{12}CO_3^- \leftrightarrow {}^{12}CO_{2\,(aq)} + H^{13}CO_3^- \quad (11)$$

is fast on the timescale of the $CO_2$ reduction reaction (sufficiently fast that it alters the isotopic composition of $CO_{2\,(aq)}$ and $CO_{2\,(gas)}$ in a different manner). This observation provides further support for our mechanism, along with the idea that $CO_2H$ behaves like $HCO_3^-$ or $H_2CO_3$ when attached to the catalyst.

Lastly, we remark that our proposed mechanism gives a prediction regarding the kinetics of the reaction that may be verified experimentally. Assuming that the catalyst-$CO_2$ complex exists mostly as $[Co^IN_4H]^+-CO_2H$ (**6**) under reaction conditions, the rate of CO formation should increase linearly with the partial pressure of $CO_2$. To the best of our knowledge, the dependence of the rate of the reaction on the partial pressure of $CO_2$ has not been reported in the literature. Thus, we suggest that investigating this dependence is a well grounded way to probe the mechanism of $CO_2$ reduction catalyzed by **1**.

## IV. CONCLUSIONS

Based on DFT calculations, we deduced a mechanism for the selective reduction of $CO_2$ to CO by the tetraaza $[Co^{II}N_4H]^{2+}$ catalyst (**1**). Our simulations provide structural insight on the reasons for the catalyst's selectivity: hydrogen bonding between bound $CO_2$ and the NH group in the tetraaza ring is vital. Analysis of our results confirms previous speculation (from experimental observations) regarding the role of $CO_2$ acting as a Lewis acid to assist in the dissociation of the C−O bond to generate $HCO_3^-$ and CO. This step is rate-determining and the calculated TOF of $1 \times 10^{-3}$ s$^{-1}$ is in agreement with the experimental value of $1^{-2}$ s$^{-1}$ (in terms of energy, less than 2 kcal/mol difference in effective $\Delta G^{\ddagger}$ for the catalytic cycle). No other $CO_2$ reduction reaction pathway was found to be competitive: the activation energies for alternative paths are higher than the "CO + $HCO_3^-$" path by a large ($> 10$ kcal/mol) margin—much larger than the typical DFT error range ($\approx 4$ kcal/mol). Considering this in tandem with the aforementioned experimental reports from the literature, does not leave much room for other possible $CO_2$ reduction pathways. Finally, the mechanism makes a prediction that may be experimentally verified: that the rate of CO formation should increase linearly with the partial pressure of $CO_2$.

## V. ACKNOWLEDGEMENTS

A.J.G, A.T.B., and M.H.G acknowlege the Joint Center for Artificial Photosynthesis, a DOE Energy Innovation Hub, supported through the Office of Science of the U.S. Department of Energy under Award DE-SC00004993. S.P. and J.L.M-C. were supported by Florida State University (FSU). The computing work perfomed at LBNL used resources of the National Energy Research Scientific Computing Center, a DOE Office of Science User Facility supported by the Office of Science of the U.S. Department of Energy under Contract No. DE-AC02-05CH11231; the computing work at FSU was performed on the High Performance Computer cluster at the FSU Research Computing Center.